\documentclass[preprint,aps]{revtex4}

\usepackage{graphicx}
\usepackage{dcolumn}
\usepackage{bm}

\begin{document}

\preprint{}

\title{Limits to the sensitivity of a low noise compact atomic gravimeter}

\author{J. Le Gou\"{e}t}
\author{T. E. Mehlst\"{a}ubler}%
\altaffiliation[]{Physikalisch-Technische Bundesanstalt (PTB), Bundesallee 100, D-38116 Braunschweig, Germany}

\author{J. Kim}%
\altaffiliation[]{Physics Department, Myongji University, Korea.}

\author{S. Merlet}%
\author{A. Clairon}%
\author{A. Landragin}%
\author{F. Pereira Dos Santos}%
 \email{franck.pereira@obspm.fr}

\affiliation{%
LNE-SYRTE, CNRS UMR 8630, UPMC, Observatoire de Paris \\
61 avenue de l'Observatoire, 75014 Paris, France
}%

\date{\today}

\begin{abstract} A detailed analysis of the most relevant sources
of phase noise in an atomic interferometer is carried out, both
theoretically and experimentally. Even a short interrogation
time of 100 ms allows our cold atom gravimeter to reach an excellent
short term sensitivity to acceleration of $1.4\times 10^{-8}$g at
1s. This result relies on the combination of a low phase noise
laser system, efficient detection scheme and good shielding from
vibrations. In particular, we describe a simple and robust
technique of vibration compensation, which is based on correcting
the interferometer signal by using the AC acceleration signal measured
by a low noise seismometer.

\end{abstract}

\maketitle

\section{Introduction}

Over the last fifteen years, atom interferometry techniques
\cite{Borde89} have been used to develop novel inertial sensors,
which now compete with state of the art ``classical''
instruments~\cite{Niebauer95}. After the first demonstration
experiments in the early 90's \cite{Kasevich91,Riehle91}, the
performance of this technology has been pushed and highly
sensitive instruments have been realized. As the inertial phase
shifts scale quadratically with the interrogation time, high
sensitivies can be reached using either cold atoms along parabolic
trajectories \cite{Kasevich91,Canuel06}, such as in microwave
fountain clocks, or very long beam machines. Best short term
sensitivities to acceleration of $0.8-1.1\times 10^{-7}\textrm{m
s}^{-2}$ at 1s , and to rotations of $6\times 10^{-10}\textrm{rad
s}^{-1}$ at 1s have been reached in the experiments developed in
the groups of S. Chu \cite{Peters01} and M. Kasevich
\cite{Gustavson02}. Moreover, a key feature of these instruments
is to provide an absolute measurement with improved long term
stability compared to other sensors, due to the intrinsic stability of their
scale factor and a good control of the environment of the atomic samples.

Applications of this kind of inertial interferometers are growing, from the measurement
of fundamental constants, such as the gravitational constant G
\cite{Bertoldi06,Fixler07}, to the development of transportable
devices for navigation, gravity field mapping, detection of
underground structures and finally for space missions \cite{APB},
where ultimate performances can be met, thanks to the absence of
gravity and a low vibration environment.

We are currently developing a cold atom gravimeter, within the
frame of the watt balance project, conducted by the Laboratoire
National de M\'{e}trologie et d'Essais (LNE)~\cite{Geneves05}. A
watt balance allows to link the unit of mass, the kilogram, to
electrical units and provide a measurement of the Planck constant.
Watt Balances developed at NIST and NPL presently reach relative
accuracies of a few parts in $10^8$ \cite{Steiner07,Robinson07}.
During one of the phases of this experiment, the weight of a test
mass is balanced with an electric force. An absolute measurement
of gravity experienced by the test mass is thus required, which
will be realized with our atom interferometer with a targeted
relative accuracy of 1 ppb.

In this paper, we describe the realization of this sensor, which
has been designed to be relatively compact, in order to be easily
transportable. Our gravimeter reaches, despite rather small
interaction times, a sensitivity of $1.4\times 10^{-8}$g at 1s,
better than state of the art ``classical''
gravimeters, and comparable to much larger atomic
fountain gravimeters \cite{Peters01}.

In this paper, we first describe our experimental setup,
and we investigate in detail in the next sections the
contributions of the different sources of noise which affect the
sensitivity. In particular, we describe in section \ref{vib} an
original, simple and efficient technique of vibration
compensation, which allows to improve the sensitivity of our
measurement by rejecting residual vibrational noise with the help
of a low-noise seismometer \cite{Yver03}.

\section{Experimental setup}
\label{expsetup}

The principle of the gravimeter is based on the coherent splitting
of matter-waves by the use of two-photon Raman transitions \cite{Kasevich91}. These
transitions couple the two hyperfine levels $F=1$ and $F=2$ of the
$^5S_{1/2}$ ground state of the $^{87}$Rb atom. An intense beam of
slow atoms is first produced by a 2D-MOT. Out of this beam $10^7$
atoms are loaded within 50 ms into a 3D-MOT and subsequently
cooled in a far detuned (-25 $\Gamma$) optical molasses. The
lasers are then switched off adiabatically to release the atoms
into free fall at a final temperature of $2.5~\mu\textrm{K}$. Both
lasers used for cooling and repumping are then detuned from the
atomic transitions by about 1~GHz to generate the two off-resonant
Raman beams. For this purpose, we have developed a compact and agile laser
system that allows us to rapidly change the operating frequencies
of these lasers, as described in \cite{Cheinet06}. Before entering
the interferometer, atoms are selected in a narrow vertical velocity
distribution ($\sigma_v \leq v_r = 5.9$ {mm/s}) in the $\left|F=1,
m_F=0\right\rangle$ state, using a combination of microwave and
optical Raman pulses.

The interferometer is created by using a sequence of three pulses
($\pi/2-\pi-\pi/2$), which split, redirect and recombine the
atomic wave packets. Thanks to the relationship between external
and internal state \cite{Borde89}, the interferometer phase shift
can easily be deduced from a fluorescence measurement of the
populations of each of the two states. Indeed, at the output of
the interferometer, the transition probability $P$ from one hyperfine
state to the other is given by the well-known relation for a two
wave interferometer : $P=\frac{1}{2}\left(1 +
C\cos\Delta\Phi\right)$, where $C$ is the interferometer contrast,
and $\Delta\Phi$ the difference of the atomic phases accumulated
along the two paths.  The difference in the phases accumulated
along the two paths depends on the acceleration $\vec{a}$
experienced by the atoms. It can be written as
$\Delta\Phi=\phi(0)-2\phi(T)+\phi(2T)=-\vec{k}_{eff} \cdot
\vec{a}T^{2}$ \cite{Borde01}, where $\phi(0,T,2T)$ is the difference of the
phases of the lasers, at the location of the center of the atomic
wavepackets, for each of the three pulses. Here
$\vec{k}_{eff}=\vec{k}_{1}-\vec{k}_{2}$ is the effective wave
vector (with $|\vec{k}_{eff}|=k_1+k_2$ for counter-propagating
beams), and $T$ is the time interval between two consecutive
pulses.

The Raman light sources are two extended cavity diode lasers based
on the design of \cite{Baillard06}, which are amplified by two
independent tapered amplifiers. Their frequency difference is
phase locked onto a low phase noise microwave reference source. The two
Raman laser beams are overlapped with a polarization beam splitter
cube, resulting in two orthogonal polarized beams. First, a small
part of the overlapped beams is sent onto a fast photodetector to
measure the optical beat. This beat-note is mixed down with the
reference microwave oscillator, and compared to a stable reference
RF frequency in a Digital Phase Frequency Detector. The phase
error signal is then used to lock the laser phase difference at
the very position where the beat is recorded. The phase lock loop
reacts onto the supply current of one of the two lasers (the
``slave'' laser), as well as on the piezo-electric transducer that
controls the length of its extended cavity. Finally, the two
overlapped beams are injected in a polarization maintaining fiber,
and guided towards the vacuum chamber. We obtain
counter-propagating beams by placing a mirror and a quarterwave
plate at the bottom of the experiment. Four beams are actually
sent onto the atoms, out of which only two will drive the
counter-propagating Raman transitions, due to conservation of
angular momentum and the Doppler shift induced by the free fall of
the atoms.

\begin{figure}
     \includegraphics[width=9cm,angle=-90]{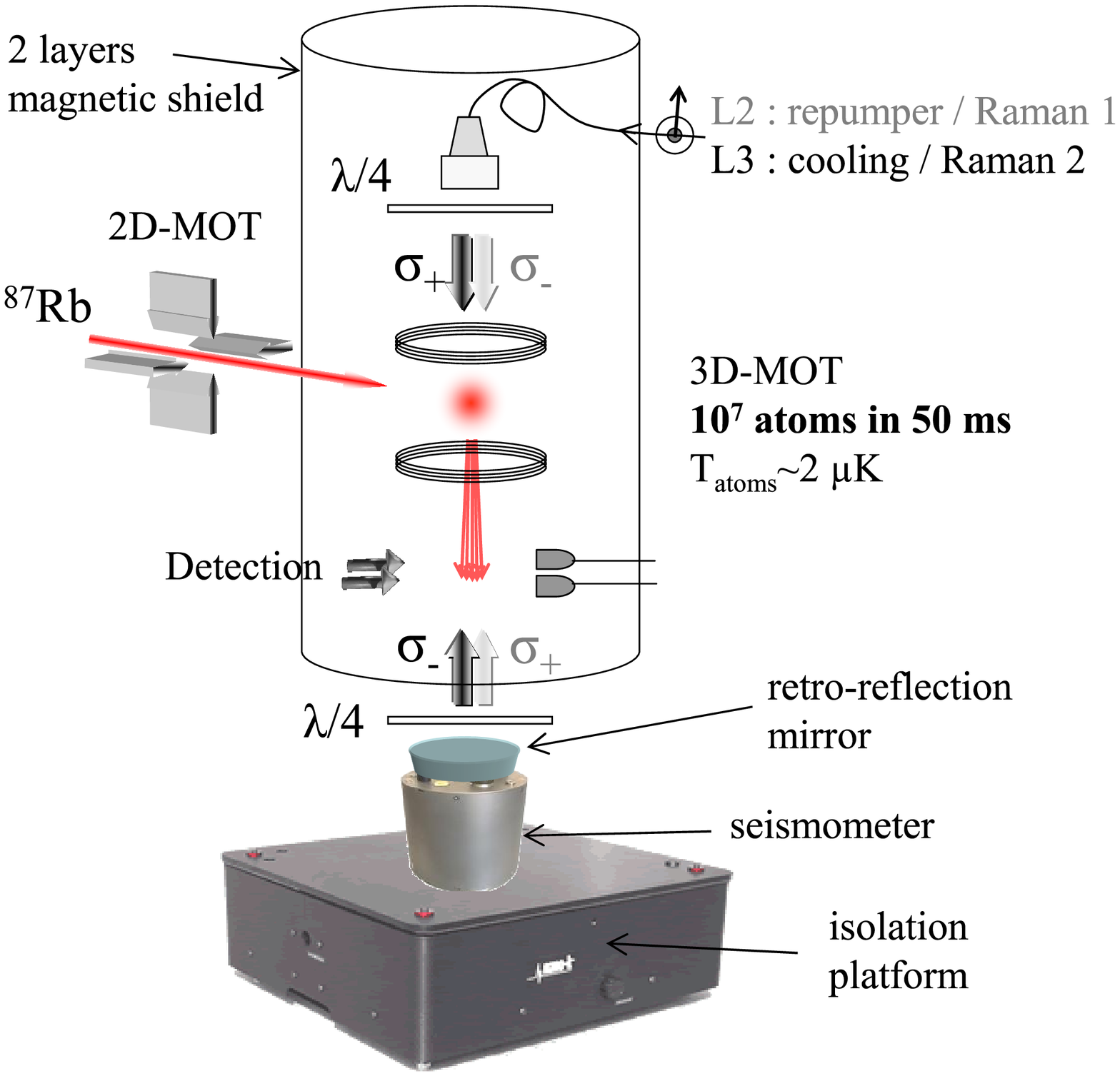}
    \caption{Scheme of the experimental set-up. The interferometer is realized in a stainless steel vacuum chamber, shielded form magnetic field fluctuations with two layers of mu metal. The chamber is sustained with three legs onto a passive isolation platform (Minus-K BM). Atoms are first trapped in a MOT, cooled with optical molasses and released. The interferometer is then realized during their free fall, with vertical Raman laser beams, which enter the vacuum chamber from the top, and are retro-reflected on a mirror located below. This mirror is attached to a low noise seismometer. Finally, the populations in the two hyperfine states are measured by fluorescence, which allows to determine the interferometer phase shift.}
    \label{}
\end{figure}

\section{Short term sensitivity}

\subsection{Detection Noise}

The transition probability is deduced from the population in each
of the two hyperfine states, which are measured by fluorescence.
Ultimately, noise on this measurement is limited by the quantum
projection noise $\sigma_{P}=1/2\sqrt{N}$ \cite{Itano93}, where $\sigma_{P}$ is the standard deviation of the transition probability, and $N$ the number of detected atoms. Other sources of
noise, such as electronic noise in the photodiodes, laser
frequency and intensity noise, will affect the measurement, and
might exceed the quantum limit depending on the number of atoms \cite{Santarelli99}.

\subsubsection{Basic scheme}

The detection system implemented at first in the gravimeter is
similar to the one developed for atomic fountain clocks. It
consists of two separated horizontal sheets of light. The first
detection zone consists in a laser beam circularly polarized,
tuned on the cycling transition ($F=2 \rightarrow F=3$). This beam
is retro-reflected on a mirror, in order to generate a standing
wave. It allows to measure  the number of atoms in the $F=2$ state
by fluorescence. The atoms which have interacted with the laser
light are then removed by a pusher beam, obtained by blocking the
lower part of the retro-reflected beam. The second zone is a
repumper beam tuned on the ($F=1 \rightarrow F=2$), which pumps
the atoms in $F=1$ into $F=2$. The last zone is a standing wave
tuned on the cycling transition, which allows to measure the
number of atoms initially in $F=1$. The fluorescence emitted in
the upper and lower zones is detected by two distinct photodiodes
that collect 1 \% of the total fluorescence. Using a $\pi/2$ microwave pulse,
the noise on the measurement of the transition probability at mid-fringe ($P=0.5$)
has been measured in the gravimeter as a function of the number of
atoms, see figure~\ref{Detection_noise}. The saturation parameter
at the center of the detection beams is close to 1. For less than
$10^6$ atoms, the detection noise is limited by the noise of the
photodiodes and the electronics. The standard deviation of the
fluctuations of the transition probability $\sigma_P$ is then
inversely proportional to the number of atoms. The detection noise
is equivalent to about 900 atoms per detection zone. For atom
number larger than $5\times10^6$, technical noise, arising from
intensity and frequency noise of the detection laser, limits
$\sigma_P$ to about $3\times 10^{-4}$. The noise in the measured
transition probability $\sigma_P$ converts into phase noise
$\sigma_{\phi}= 2/C \times \sigma_P$, with $C$ being the contrast
of the interferometer. A sensitivity close to 1 mrad per shot for
the interferometer measurement can thus be obtained for the
interferometer when the number of atoms is larger than $5\times
10^6$.

\subsubsection{Improved scheme}

In our experiment, the same laser system is used for the Raman
interferometer and for the atom trapping. After the interferometer
sequence the lasers are brought back close to resonance in order
to trap the atoms during the beginning of the next experimental
cycle. In principle, they can also be used to detect the atoms by
pulsing the vertical beam, when the atoms are located in the
detection region.

\begin{figure}
        \includegraphics[width=9 cm]{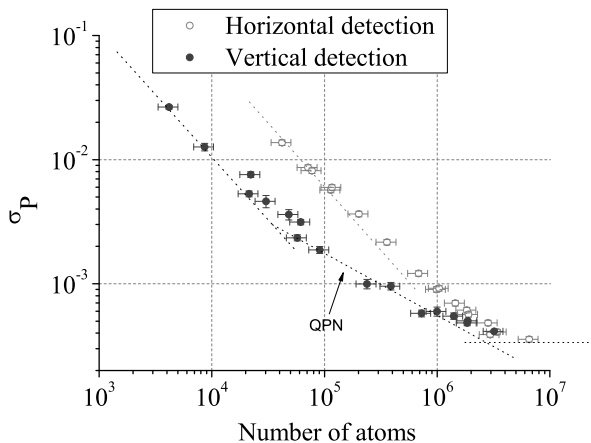}
    \caption{Allan standard deviation of the excitation probability $\sigma_P$ for both methods described. Open circles correspond
    to horizontal detection, full circles to vertical detection. QPN is the Quantum projection Noise limit. The phase noise introduced onto the
    atom interferometer is $\sigma_{\phi}= 2/C \times \sigma_P$.}
    \label{Detection_noise}
\end{figure}

The detection sequence we use
has been inspired by the detection system \cite{McGuirk01} of the gradiometer of
\cite{McGuirk02}. (1)  When the atoms are located in front of the
top photodiode, a low intensity pulse, slightly red detuned to the
cycling transition is induced in order to stop the $F=2$ atoms.
(2) One then waits for the  atoms in $F=1$ to reach the position
in front of the bottom photodiode. (3)  A second pulse (10 ms
long) is then applied at full power. Cooling and repumper beams
are both present during this pulse. (4)  A third pulse,  10 ms
later, finally serves for background substraction of stray light.
The areas of the fluorescence signals collected by the two
photodiodes are thus proportional to the number of atoms in each
of the two hyperfine states.
This detection scheme has several advantages. First, the intensity in the
vertical beam is much higher than in the standard detection beam,
the saturation parameter being close to 50. Atoms will thus remain resonant despite the heating induced by photon recoils. Second, atoms spend a longer time in front of the photodiodes. Finally, repumper is present on both clouds during the whole duration of the detection pulse.
With this scheme, the detection noise is now equivalent to only 150 atoms per zone, thanks to the increase of the fluorescence signal, see figure~\ref{Detection_noise}. The detection is at the quantum projection noise limit with about $10^5$ atoms. The same limit of $\sigma_P \approx 3\times 10^{-4}$ is found for large
number of atoms. This detection scheme is thus more efficient for low
number of atoms, for instance when using Bose Einstein condensates or
narrower velocity selection. Limits to the signal to noise ratio
for large numbers of atoms still have to be identified. They could
arise from laser intensity and frequency fluctuations, as well as
fluctuations of the normalization. In principle, the second
detection scheme should be insensitive to laser fluctuations,
which are common mode for the two populations, as the measurements
are performed simultaneously.

\subsection{Phase noise}
\label{phasenoise}

In our interferometer, the noise in the phase difference of the
two Raman lasers induces fluctuations of the interferometer phase.
In a previous publication \cite{Cheinet05}, we have introduced a
useful tool to calculate the influence of the different sources of
noise onto the stability of the interferometer phase measurement,
the sensitivity function, $g(t)$~\cite{Dick87}. This function quantifies
the influence of a relative laser phase shift $\delta\phi$
occurring at time $t$ onto the phase of the interferometer $\delta
\Phi(\delta\phi,t)$. When operating the interferometer at mid-fringe, it can be written as:
\begin{equation} \label{g(t)}
g(t)=\lim_{\delta\phi\rightarrow 0}\frac{\delta
\Phi(\delta\phi,t)}{\delta\phi}.
\end{equation}
The interferometer phase shift $\Phi$ induced by fluctuations of $\phi$ is then given by:
\begin{equation}
\label{phi}
\Phi=\int^{+\infty}_{-\infty}{g(t)\frac{d\phi(t)}{dt}dt}.
\end{equation}
For completeness we briefly recall the expression of the
sensitivity function. With a sequence of three pulses
$\pi/2-\pi-\pi/2$ of duration $\tau_R - 2\tau_R - \tau_R$ and a
time origin chosen at the center of the $\pi$ pulse, $g$ is an odd
function whose expression was first derived in \cite{Cheinet05}:
\begin{equation} \label{g_entier}
g(t) = \left\{\begin{array}{ll}
            \sin \Omega_R t & \textrm{for $0<t<\tau_R$}\\
            1 & \textrm{for $\tau_R<t<T+\tau_R$}\\
            -\sin \Omega_R (T - t) & \textrm{for $T+ \tau_R<t<T+2\tau_R$}\\
            \end{array} \right.
\end{equation}
where $\Omega_R$ is the Rabi frequency.

The noise of the interferometer is characterized by the Allan variance of the interferometric phase fluctuations,
$\sigma^{2}_{\Phi}(\tau)$, defined as:
\begin{eqnarray}
 \sigma_{\Phi}^{2}(\tau)&=&\frac{1}{2}\langle(\bar{\delta \Phi}_{k+1}-\bar{\delta \Phi}_{k})^{2}\rangle \\
&=&\frac{1}{2}\lim_{n\rightarrow \infty}\left\{
 \frac{1}{n}\sum_{k=1}^{n}(\bar{\delta \Phi}_{k+1}-\bar{\delta \Phi}_{k})^{2}\right\}.\label{eq:variance_allan}
 \end{eqnarray}
Here $\bar{\delta \Phi}_{k}$ is the average value of $\delta \Phi$
over the interval $[t_{k},t_{k+1}]$ of duration $\tau$. The Allan
variance is equal, within a factor of two, to the variance of the
differences in the successive average values $\bar{\delta
\Phi}_{k}$ of the interferometric phase. Our interferometer
operates sequentially at a rate $f_c=1/T_{\rm{c}}$, where $\tau$
is a multiple of $T_c$ : $\tau=m T_c$. Without loosing generality,
we can choose $t_{k}=-T_c/2+k m T_c$.

For large averaging times $\tau$, the Allan variance of the
interferometric phase is given by \cite{Cheinet05}:
\begin{equation}
\label{Dick} \sigma^{2}_{\Phi}(\tau)={1\over \tau}\sum_{n=1}^{\infty}|H(2\pi n f_{\rm{c}})|^2
        S_{\phi}({2\pi n f_{\rm{c}}}).
\end{equation}

The transfer function is thus given by $H(\omega)=\omega
G(\omega)$, where $G(\omega)$ is the Fourier transform of the
sensitivity function:
\begin{equation}
G(\omega)= \int_{-\infty}^{+\infty}e^{-i\omega t}g(t)dt
\label{eq:G1}.
\end{equation}

Equation \ref{Dick} shows that the sensitivity of the
interferometer is limited by an aliasing phenomenon similar to the
Dick effect in atomic clocks \cite{Dick87}: only the phase noise
at multiples of the cycling frequency appears in the Allan
variance, weighted by the Fourier components of the transfer
function.

\begin{figure}
     \includegraphics[width=8cm,angle=-90]{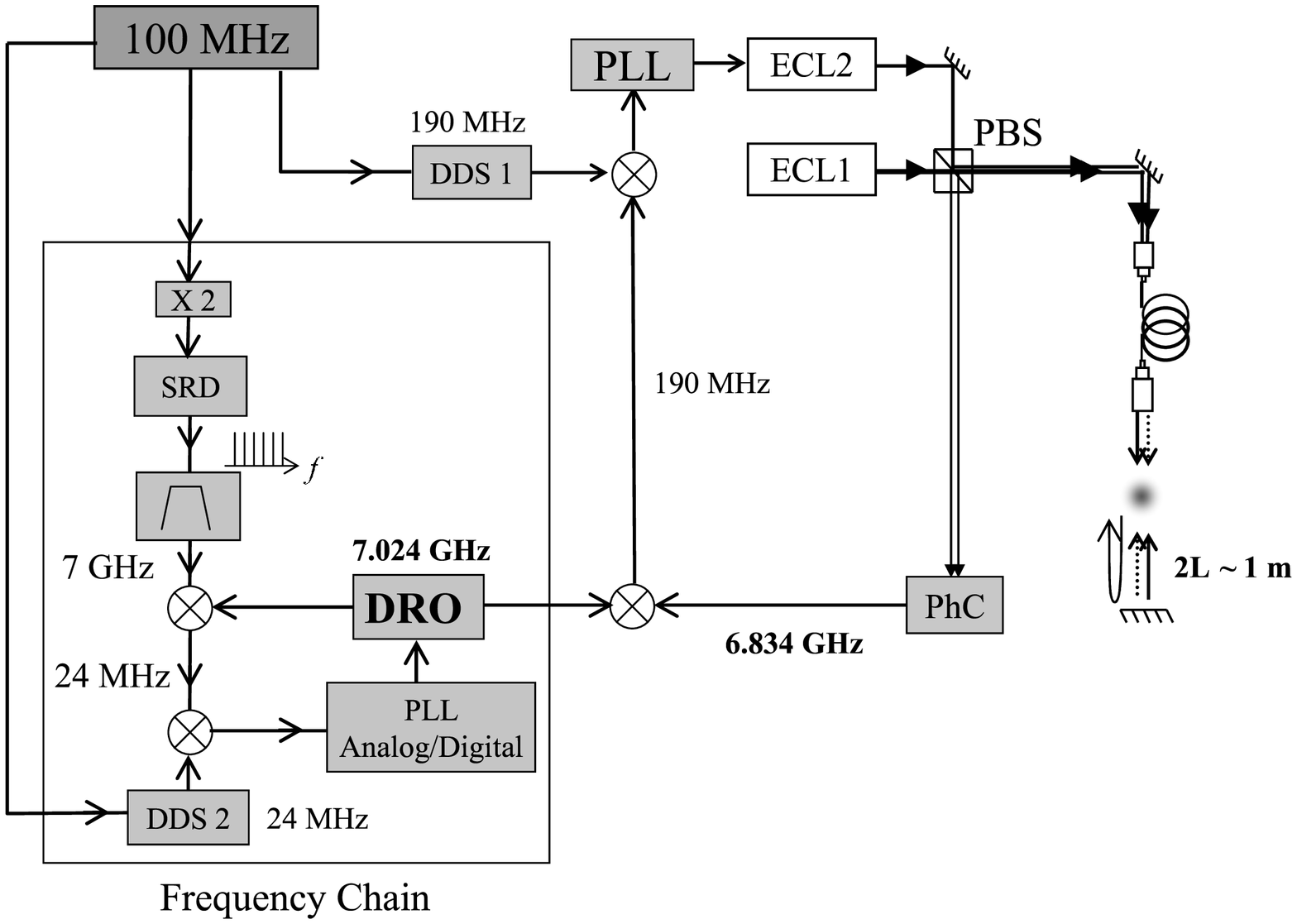}
    \caption{Scheme of the microwave synthesis and phase locked lasers.
The 100 MHz quartz signal enters a frequency chain which generates
the reference microwave frequency. Its microwave output is mixed
with the lasers beatnote, and the IF is compared with a DDS
signal. The error signal is used to phase lock the Raman lasers.
ECL Extended Cavity Laser, DDS Direct Digital Synthesizer,
PhC Photoconductor, SRD Step Recovery Diode, PBS Polarizing beam splitter, DRO Dielectric Resonator Oscillator.}
    \label{picture-MW}
\end{figure}

\subsubsection{Reference oscillator}

The previous formalism is used to analyze the specifications
required for the reference microwave frequency \cite{Cheinet05}.
We choose to generate the reference microwave signal by
multiplication of an ultra-stable quartz. Assuming perfect
multiplication of state of the art ultra-stable quartz
oscillators, we find that their phase noise at low frequency will
limit the sensitivity of the interferometer phase measurement at
the mrad per shot level, for our total interferometer time of
$2T=$ 100 ms. A noise of 1 mrad/shot corresponds to a sensitivity
to acceleration of $1.2\times10^{-9}g$ at 1~s, for a repetition rate of 4 Hz and interaction time $2T=100$ms.

Moreover, the relatively short duration of the Raman
pulses makes the interferometer particularly sensitive to high
frequency noise. If we consider a Raman pulse duration of 10
$\mu$s, a white noise floor of the reference microwave source, with a PSD of -120 dBrad$^2$/Hz,
contributes to the interferometer phase noise at the level of 1 mrad/shot.

\begin{figure}
        \includegraphics[width=9cm]{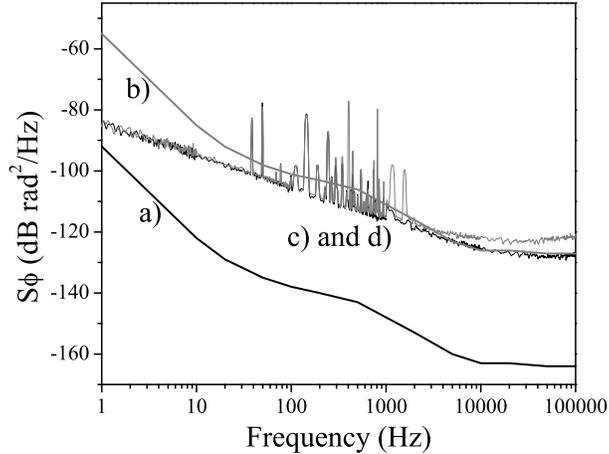}
    \caption{Power spectral density of phase fluctuations of the reference 100 MHz oscillator
at 100 MHz (trace a) and at 6.8 GHz (trace b) assuming no
degradation. c) and d) display the power spectral density of the
phase noise generated by the synthesis, with a digital PLL
(trace c, in grey) and analog PLL (trace d, in black)}
    \label{Quartzchain}
\end{figure}

The required phase noise specifications
for such a reference oscillator cannot be met both at low and high
frequency by a single quartz, but can be achieved by phase locking
two quartz oscillators. We use a combination of two quartz : a
Ultra Low Noise 10 MHz quartz (Blue Top from Wenzel) is
multiplied up to 100 MHz, to which a 100 MHz SC Premium (Wenzel) is phase locked with
a bandwidth of about 400 Hz. This system, whose performance is
indicated as trace (a)t in figure \ref{Quartzchain}, was realized
by Spectradynamics Inc. Measurements performed in our laboratory
on two independent such systems confirmed these specifications. If
this oscillator is multiplied to 6.8 GHz without any degradation -
see trace (b) in figure \ref{Quartzchain} - we calculate that its
phase noise degrades the sensitivity of the interferometer at the
level of 1.2 mrad per shot.

\subsubsection{Microwave frequency synthesis}

The microwave chain generates the 6.834 GHz reference signal, used to
lock the laser to the Raman transition, out of a stable reference
quartz (see figure~\ref{picture-MW}). Details on this synthesis and
its performance have been published in a previous publication
\cite{Nyman06}. We briefly recall its architecture in the
following paragraph.

In a first stage the 100 MHz output of the quartz system is
multiplied by 2. Then, the 200 MHz output is amplified and sent to
a Step Recovery Diode, which generates a comb of multiples of 200
MHz. The 35th harmonic is selected with a passive filter, and
compared in a mixer with a Dielectric Resonator Oscillator (DRO)
operating at 7.024 GHz. The 24 MHz intermediate frequency is mixed
again with a Direct Digital Synthesizer (DDS) using a digital
phase/frequency detector. Using the phase error signal, the DRO is
finally phase locked onto the comb with an offset frequency
controlled by the DDS (DDS2 on figure \ref{picture-MW}). Figure
\ref{Quartzchain} diplays the phase noise power spectral density
of the microwave chain, which has been measured by comparing the
outputs of two identical chains, that shared a common 100 MHz
input. The degradation generated by the system on the phase noise
has been measured. It contributes to 0.6 mrad/shot to the
sensitivity of the interferometer measurement. Figure
\ref{Quartzchain} also displays the phase noise obtained when
replacing the Digital phase detector by an analog mixer, which
allows to reach a lower white phase noise floor at high frequency.
Still, we currently use the digital phase detector, as the lock
loop is then more robust, the DRO stays locked even if the DDS
frequency is changed rapidly by several MHz.

\subsubsection{Laser phase lock}

\begin{figure}
        \includegraphics[width=9cm]{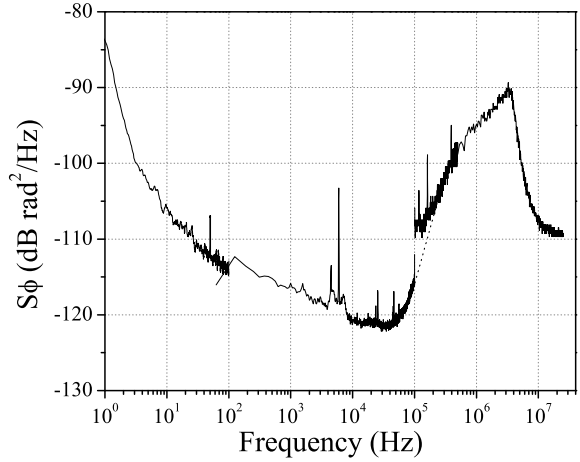}
    \caption{Power spectral density of laser phase noise. The measurement is performed
by comparing the laser phase with respect to the reference
oscillator, with a FFT analyzer below 100 kHz, and with a spectrum
analyzer above. Note that the intrinsic phase noise of our
spectrum analyzer is about -110 dBrad$^2$/Hz at 100 kHz, which is
above the lasers phase noise. The bandwidth of the PLL is 3.5
MHz.}
    \label{PhaseLock}
\end{figure}

The phase difference between the two Raman lasers is locked onto
the phase of the reference microwave signal with an electronic phase lock loop
(PLL) \cite{Santarelli94}. We have experimentally measured  the residual phase noise
power spectral density of our phase lock system. The measurement
was performed by mixing the intermediate signal at 190 MHz and the
local oscillator DDS1 onto an independent RF mixer, whose output
phase fluctuations were analyzed with a FFT analyzer (for
frequencies less than 100 kHz) and a RF spectrum analyzer (for
frequencies above 100 kHz). The result of the measurement is
displayed in figure \ref{PhaseLock}. The phase noise decreases at
low frequencies down to a minimum value of -121~dBrad$^2/$Hz at
about 30 kHz. At this frequency, we found that the residual noise
was not limited by the finite gain of the PLL, but by the
intrinsic noise of the PLL circuit. Above 60 kHz, the noise
increases up to -90~dBrad$^2/$Hz at 3.5 MHz, which is the
natural frequency of our servo loop. The contribution of the residual
noise is dominated by this high frequency part. We calculate a
contribution to the interferometer phase noise of 1.5 mrad/shot.

\subsubsection{Propagation in  the fiber}
\begin{figure}
     \includegraphics[width=9cm]{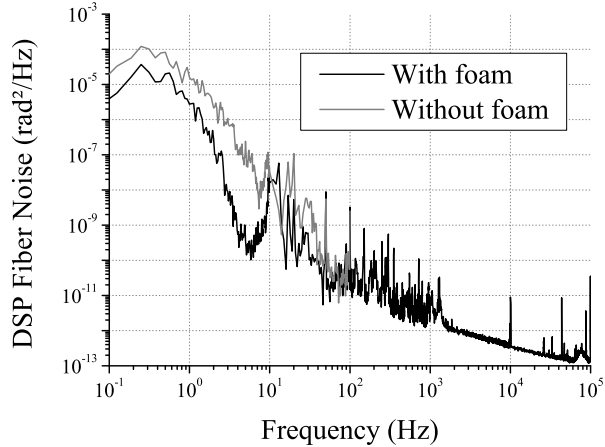}
\caption{Power spectral density of phase fluctuations induced by
the propagation in the fiber, with foam (black trace) and without
(grey foam). The two Raman lasers travel with two orthogonal
polarizations in the same polarization maintaining fiber. The
noise, dominated by low frequency noise due to acoustic noise and
thermal fluctuations, gets significantly reduced by shielding the
fiber with foam.}
     \label{DSPFibre}
\end{figure}

In our experiment, the Raman beams are generated by two
independent laser sources. The beams are finally overlapped by
mixing them on a polarizing beam splitter cube, so the beams have
orthogonal polarizations. A small fraction of the total power is
sent to one of the two exit ports of the cube, where a fast
photodetector detects the beat frequency. The beat note is
compared with the reference signal produced by the microwave
chain, in order to phase lock the lasers. The laser beams, sent to
the atoms, are diffracted through an acousto-optical modulator, used as an optical shutter to produce the Raman pulses.
During the interferometer, the total power is diffracted in the
first order to produce the vertical Raman beams. Both beams are
guided towards the atoms with a polarization maintaining fiber.
Since the Raman beams have orthogonal polarization, any fiber
length fluctuation will induce phase fluctuations, due to its birefringence. We measured the phase noise induced
by the propagation in the fiber by comparing the beat signal
measured after the fiber with the one we use for the phase lock.
Figure \ref{DSPFibre} displays the power spectral density of the
phase noise induced by the propagation, which is dominated by low
frequency noise due to acoustic noise and thermal fluctuations.
This source of noise was reduced by shielding the fiber from the
air flow of the air conditioning, surrounding it with some
packaging foam. The calculated contribution to the interferometer phase noise is 1.0
mrad/shot. This source of noise can be suppressed by using two
identical linear polarizations for the Raman beams.

\subsubsection{Retro-reflection delay}

The phase lock loop guaranties the stability of the phase
difference between the two Raman lasers at the particular position,
where their beatnote is measured on the fast photodetector.
Between this position and the atoms, this phase difference is
affected by fluctuations of the respective paths of the two beams
over the propagation distance. In our experiment, the influence of
path length variations is minimized by overlapping the two beams,
and making them propagate as long as possible over the same path.
However, for the interferometer to be sensitive to inertial
forces, the two beams need to be counter-propagating. The two
overlapped beams are thus directed to the atoms and finally
retro-reflected on a mirror. As a consequence, the reflected beam is delayed with respect to
the other one. The phase difference at the atoms position is then
affected by the phase noise of the lasers accumulated during this
reflection delay. This effect has been described in detail in
\cite{LeGouet07}, where the influence of the frequency noise of
the Raman lasers onto the interferometer phase was studied
quantitatively. From the measurement of the power spectral density
of the laser frequency fluctuations, we derived a contribution of
this effect of 2.0 mrad/shot. This effect can be significantly
reduced by reducing the linewidth of the reference laser,
and/or reducing the delay by bringing the mirror closer to the
atoms.

\subsubsection{Overall laser phase noise contribution}

Adding all the laser phase noise contributions described above,
leads to a sensitivity of 3 mrad/shot, which corresponds to an
acceleration sensitivity of $3.7\times10^{-9}$ g/Hz$^{1/2}$, for our interrogation time of $2T=100$ ms.

\subsection{Short term fluctuations of frequency dependent shifts}

Due to its intrinsic symmetry, the phase of the interferometer is
not sensitive to a shift of the resonance frequency $\delta\nu$ as
long as this shift is constant. On the contrary, a time dependent
frequency shift will in general lead to a phase shift of the
interferometer. For instance, sensitivity to acceleration can be
seen as arising from a time dependent Doppler shift. Such
frequency shifts are also caused by Stark shifts, Zeeman shifts,
cold atom interactions. The formalism of the sensitivity function
can be used to determine the influence of these effects \cite{TheseBen}, as
equation (\ref{phi}) can be written as
\begin{equation}
\label{phi2} \Phi=\int^{+\infty}_{-\infty}{g(t)2\pi\delta\nu(t)dt}
\end{equation}
In this section we detail  the effects of the dominant
contributions : AC Stark shifts and quadratic Zeeman shift.

\subsubsection{Light shifts}
Each of the Raman lasers, as they are detuned with respect to the
electronic transition  $5S_{1/2}-6P_{3/2}$, induces a light shift
on the two-photon Raman transition. This one-photon light shift
(OPLS) can be expressed as a linear combination of the laser
intensities, $\delta\nu=\alpha I_1+\beta I_2$. For a detuning
$\Delta\nu$ smaller than the hyperfine transition frequency, the
OPLS can be cancelled by adjusting the ratio between the two laser
beams : $\alpha I_{10}+\beta I_{20}=0$ \cite{Weiss94}. Still, intensity
fluctuations occurring on times scales shorter than the
interferometer duration $2T$ can lead to noise in the
interferometer phase, given by
\begin{equation}
\label{phils} \Phi=\int^{+\infty}_{-\infty}{g(t)h(t)2\pi(\alpha
I_1(t)+\beta I_2(t))dt},
\end{equation}
where h(t) = 1, during the Raman pulses, and 0 during free
evolution times.

Following the same formalism as used in~\cite{Cheinet05}, the
degradation of the sensitivity can be expressed as
\begin{equation}
\label{Dickls} \sigma^{2}_{\Phi}(\tau)={1\over \tau}(2\pi\alpha
I_{10})^2\sum_{n=1}^{\infty}|G'(2\pi n f_{\rm{c}})|^2
        (S_{I1/I1_0}({2\pi n f_{\rm{c}}})+S_{I2/I2_0}({2\pi n
        f_{\rm{c}}})),
\end{equation}
where $G'$ is the Fourier transform of $g(t) \cdot h(t)$ and
$S_{Ii/Ii_0}$ is the power spectral density of relative intensity
fluctuations of the i-th laser \cite{TheseBen}.

\begin{figure}
\includegraphics[width=9 cm]{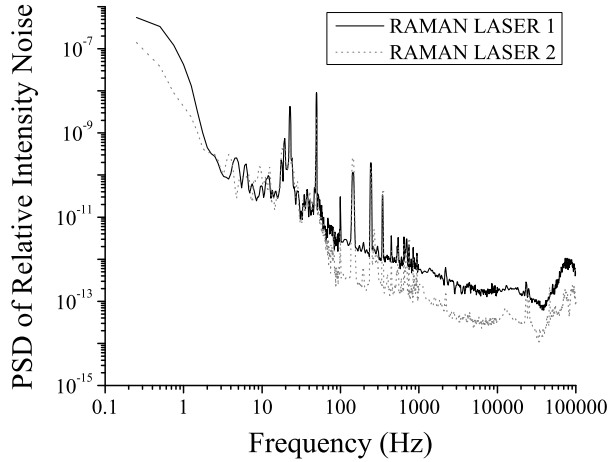}
\caption{Power spectral density of relative intensity noise (RIN)
of the Raman lasers measured at the output of the fiber.}
\label{intnoise}
\end{figure}

We measured the relative intensity noise (RIN) of both laser beams
at the output of the optical fiber. The power spectral densities
of the lasers RINs are displayed in figure \ref{intnoise}. From the measurement of the resonance condition as a function of laser intensities, we determined $\alpha I_{10}=70$ kHz, from which we finally calculate a
contribution of 0.8 mrad per shot.

In the geometry of our experiment, where a pair of co-propagating
beams is retro-reflected in order to generate the
counter-propagating laser beams, another frequency shift occurs
due to non-resonant two photons transitions. The main contribution
arises from the two-photons transition with inverted $k_{eff}$,
additional contributions are also present from co-propagating and magnetic
field sensitive transitions, $\Delta m=2$. This two photons shift,
as detailed in~\cite{TheseBen} and \cite{Clade}, can be expressed as
\begin{equation}
\label{phitp}
\Phi_{TPLS}=\int^{+\infty}_{-\infty}{g(t)h(t)\sum_{i}^{}\frac{\hbar
\Omega_i^2}{4\delta_i}}\, ,
\end{equation}
where $\Omega_i$ and $\delta_i$ are the Rabi frequency and
detuning with respect to the two photons transition. Here again,
fluctuations in the intensities will induce noise on the
interferometer phase. As $\delta_i$ increases with time due to the
increasing Doppler shift, the influence of the second and third
Raman pulse can be neglected, and eq.~\ref{phitp} approximated by
\begin{equation}
\label{phitp2} \Phi_{TPLS}=\frac{1}{\Omega}\sum_{i}^{}\frac{\hbar
\Omega_i^2}{4\delta_i} \, .
\end{equation}
As $\Omega_i,\Omega \propto \sqrt{I_1I_2}$,  this leads to
\begin{equation}
\label{phitp3} \delta\Phi_{TPLS}=\Phi_{TPLS}(\frac{\delta I_1}{2
I_1}+\frac{\delta I_2}{2 I_2}) \, .
\end{equation}

As this effect scales linearly with the Rabi frequency, it is measured with a differential measurement, alternating measurements with two different Rabi frequencies. We find $\Phi_{TPLS}=$ 40 mrad. Shot to shot fluctuations
of the relative intensity are measured to be $3 \times 10^{-4}$. The
contribution is thus about 0.1 mrad/shot.

\subsubsection{Magnetic fields}
In order to reduce the sensitivity to magnetic field fluctuations,
the atoms are selected in the $m_F=0$ state. Still, the Raman
resonance condition exhibits a quadratic Zeeman shift of
$\delta\nu=K B^2$ where $B$ is the amplitude of the magnetic field
and $K=575 $Hz/G$^2$. Magnetic field gradients will thus induce a
shift of the interferometer phase given by
\begin{equation}
\label{sphiB} \Phi=\int^{+\infty}_{-\infty}{g(t)2\pi K B(t)^2dt}
\, .
\end{equation}
In our experiment we observe a large magnetic field gradient
induced by residual magnetization of the vacuum
chamber, which is made out of stainless steel. It causes a
constant phase shift of 320 mrad in our interferometer, which we
reject at a level of 1 per 300 by consecutive differential
measurements with reversed $k_{eff}$-vectors. For it's influence
on the short term stability, we determine the stability of the
magnetic fields by recording fluctuations of the resonance
condition of a field-sensitive transition. The relative stability
of the field is $10^{-4}$ per shot, which induces a negligible
phase shift fluctuation of $3\times 10^{-2}$ mrad per shot.

\subsection{Vibrations}
\label{vib}

As the interferometer phase shift is a measurement of the relative
acceleration between free falling atoms and the ``optical ruler''
attached to the phase planes of the Raman lasers, vibrations of
the experimental setup will add noise to the measurement. With the
retro-reflected geometry, the phase difference between the laser
beams depends only on the position of a single element, the
retro-reflecting mirror. For optimal performances, it is thus
mandatory to shield this element from external vibrational noise.

The degradation of the sensitivity due to parasitic vibrations can easily be derived from equation \ref{Dick}, by replacing $S_{\phi}(\omega)$ by $k_{eff}^2S_{z}(\omega)=k_{eff}^2\frac{S_{a}(\omega)}{\omega^4}$, where $S_z$ and $S_a$ are power spectral densities of position and acceleration fluctuations. The transfer function of the interferometer thus acts as a second order low pass filter, which reduces drastically the influence of high frequency noise. The sensitivity of the interferometer is finally given by:
\begin{equation}
\label{sigmavib} \sigma^{2}_{\Phi}(\tau)={k_{eff}^2\over \tau}\sum_{n=1}^{\infty}\frac{|H(2\pi n f_{\rm{c}})|^2}{(2\pi n f_{\rm{c}})^4}
        S_{a}({2\pi n f_{\rm{c}}}).
\end{equation}

\subsubsection{Isolation of the vibration noise}

\begin{figure}
        \includegraphics[width=9cm]{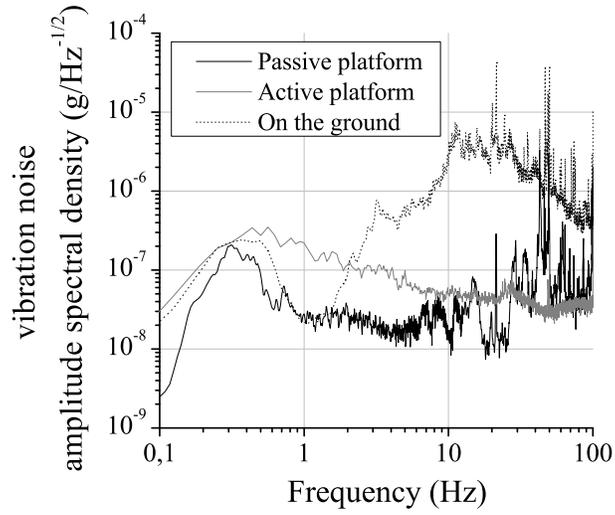}
    \caption{Comparison of the performance of two isolation platforms.
We found the passive platform behaves better in the low frequency
range, where our interferometer is most sensitive to vibrations.}
    \label{Plates_Comparison}
\end{figure}

We tested two different vibration isolation tables and compared
their performances. The tables were loaded up to their nominal
load with lead bricks. Acceleration noise on the platforms were
measured with a low noise seismometer Guralp T40. Figure
\ref{Plates_Comparison} displays the PSD of the residual vibration
noise, compared to the noise measured directly on  the ground. We
finally selected the passive platform (Minus K BM), which displays a better noise
between 0.5 and 50 Hz, where our interferometer is most sensitive.
The experimental setup was then assembled on it. With respect to the spectrum of figure \ref{Plates_Comparison}, the vibration noise of the experiment
increased at frequencies higher than 50 Hz, due to several structural
resonances excited by acoustic noise. We therefore enclosed the
experimental setup with a wooden box, whose walls were covered
with dense isolation acoustic foam. The gain on the vibration
power spectrum was about 20 dB above 50 Hz. The contribution of
residual vibrations to the interferometer can be calculated by
weighting the vibration spectrum with the transfer function of the
interferometer \cite{Cheinet05}. We then foresee a sensitivity of
$6.5\times10^{-8}$g at 1~s. Vibrations are typically less
during the night (from 1 to 5 AM) when there is no underground
traffic, and the sensitivity is then $5\times10^{-8}
$g at 1 s. Measurements of the interferometer phase noise are
in excellent agreement with the inferred vibration noise, which
surpasses all other sources of phase noise.

\subsubsection{Seismometer correction}

\begin{figure}
     \includegraphics[width=9 cm]{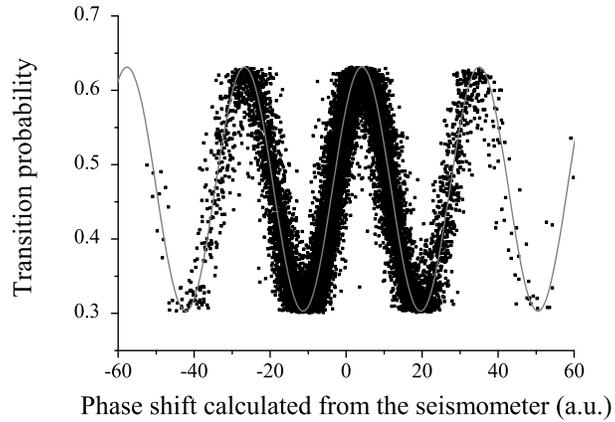}
    \caption{Interferometer signal as a function of the phase shift calculated from the seismometer signal with the passive isolation platform down. Experimental determination of the transition probabilities are displayed as dots. The line displays a sinusoidal fit to the data, with a constrained amplitude. }
    \label{Correlations}
\end{figure}

To further improve the sensitivity of the sensor, we use the
signal of the seismometer to correct the interferometer phase from
the fluctuations induced by the vibration noise. An efficient
rejection requires that the seismometer measures the vibrations of
the retro-reflecting mirror as accurately as possible. We placed
the mirror directly on top of the seismometer, which is underneath
the vacuum chamber. Figure \ref{Correlations} displays the measured transition probability as a function of the phase shift calculated from the seismometer output
signal. To obtain large variations of the interferometer phase, we
deliberately increased for this measurement the vibrational noise
by having the passive platform non-floating. This figure illustrates the good
correlation between the seismometer signal and the interferometer
phase. In the conditions of minimal noise, with a floating isolation platform, the correlation coefficient between sismometer acceleration noise and interferometer phase noise is found to be as high as 0.94.

We then studied two types of vibration compensation. The first one
is a post-correction: the velocity signal from the seismometer is
recorded during the interferometer, and we substract from the
measured interferometer phase the calculated phase shifts due to
the recorded vibrations. As the correction is applied to the
transition probability, the interferometer should be measuring at
mid-fringe, in order to have a simple (linear) relationship
between the change in the transition probability and the
fluctuation of the interferometer phase. This technique requires
that peak to peak phase noise fluctuations remain less than a few
tens of degrees and that the contrast remains constant.

\begin{figure}
     \includegraphics[width=6cm,angle=-90]{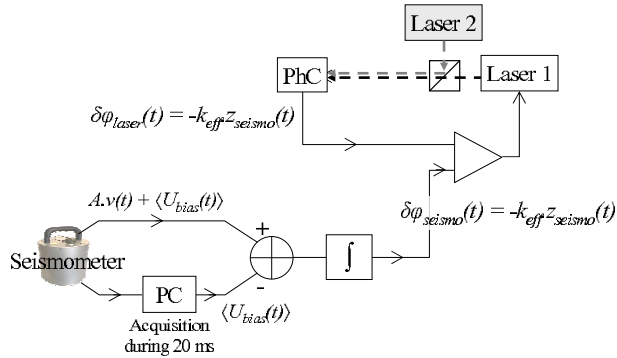}
    \caption{Scheme of the real-time compensation set-up. The phase shift induced by vibrations is subtracted from the phase error signal at the output of the phase comparator. The
laser phase difference counterbalances vibrational noise, so that
in the inertial frame, the laser phase planes are steady. PhC : photoconductor.}
    \label{LockL3Phase}
\end{figure}

The second method is a real time compensation, where the
seismometer velocity signal is amplified and integrated by an
analog circuit. The integrated signal is subtracted from the phase
error signal at the output of the comparator in the phase lock
loop of the Raman laser system. In this feed-forward loop, the
laser phase difference counterbalances vibrational noise, so that
in the inertial frame, the laser phase planes are steady. A scheme
of the setup is displayed on figure \ref{LockL3Phase}.

In principle, the digital phase detector ensures the linearity of
its voltage output with respect to the input phase difference.
Here, a maximum voltage output values $\pm 0.5 V$ corresponds to
$\pm 2\pi$ input phase difference. For the phase lock loop to
remain active, the voltage at the output of the integrator has to
correspond to a phase that remains within this range. This was
ensured by resetting the integrator at each cycle and by
compensating the intrinsic offset of the seismometer output. As
this offset fluctuates, due to either electronic noise or low
frequency vibrations, the compensation is realized by i) acquiring
the seismometer signal with a digital card at the end of the
previous measurement cycle, ii) calculating its average value, iii)
ouputting it with an analog voltage board at the beginning of the
next cycle, and iv) subtracting this last value from the seismometer
signal before being integrated. Despite all these precautions, the
range of the phase compensation had to be increased by a factor 2
(dividing the intermediate frequency signal by 2 before the phase comparator), in
order for the phase lock loop to remain active.

In principle, this last technique is more powerful, as it remains
efficient even if the contrast changes, and can compensate large
phase fluctuations, if the effective dynamic of the mixer is large
enough. In practice we found that the digital phase detector had
two drawbacks. i) The residual non-linearity of the output signal is
enough to induce a large bias to the interferometer phase (the linearity
should be at the mrad level for typical phase excursions of about 1 rad over 100 ms.)
ii) Operation far from null output increases the phase noise.
To circumvent these problems, the phase compensation could be performed in the phase lock loop of a second quartz onto the 100 MHz signal, with an analog mixer. This increases the dynamics by more than one order of magnitude and guarantees the linearity.
But, as on our platform, the vibration noise is low, we finally chose the post-correction method, which is simpler and more robust.

In both cases, the rejection efficiency was limited to a factor of
3, corresponding to a typical sensitivity of $2 \times 10^{-8}$~g
at 1s.

\subsubsection{Seismometer response function}

\begin{figure}
        \includegraphics[width=9 cm]{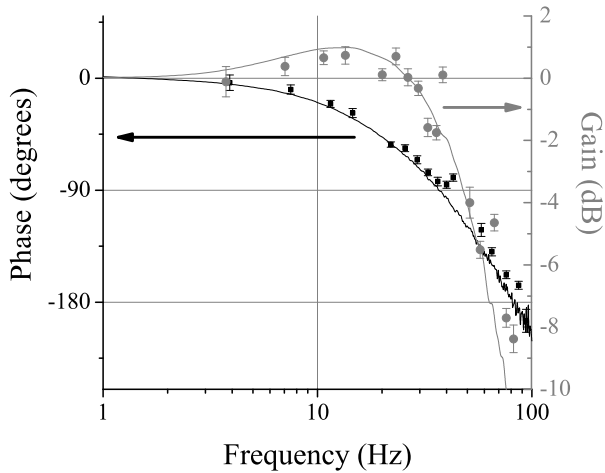}
    \caption{Transfer function between the acceleration measured
with the interferometer and the seismometer. The transfer function
agrees well with the response function of the seismometer (solid
lines).}
    \label{FT_sismo_bottom}
\end{figure}

In order to study the transfer function between the seismometer
and the retro-reflecting mirror, we induce a platform oscillation at
given frequencies (by running ac-currents through a loaded
loudspeaker placed onto the isolation platform) and record
simultaneously the atomic and seismometer signals.
The excitation
frequencies $f_{exc}$ were chosen close but not exactly equal to multiple of the cycling frequency $f_{exc}=kf_c+\delta f$, so that the transition probability and its correction calculated from the sismometer signal were modulated in time, at an apparent frequency controlled by $\delta f$. The sismometer transfer function can be determined from the ratio of the amplitudes of the modulation of the two signals and their phase difference.
The measured transfer function is displayed in figure \ref{FT_sismo_bottom}. It
corresponds very well to the response of the seismometer. This
response, which is provided by the manufacturer, can also be
retrieved by the user using an internal measurement protocol, see
solid lines in figure~\ref{FT_sismo_bottom}. As one can see from
these curves, the  seismometer has a built-in low pass filter to
cut mechanical resonances, which occur typically above 450 Hz.
This filter has a 3 dB cut-off frequency of 50 Hz. This adds some
phase shifts to the calculated correction with respect to the real
perturbation for higher frequencies. This corrupts the rejection
efficiency and eventually adds noise. To improve the rejection, we
would either need to flatten the transfer function, or to cut
frequencies at which the rejection process degrades the
sensitivity.

\subsubsection{Digital filtering}

We apply a digital filter to the seismometer signal to compensate
for attenuation and phase shifts. In the ideal situation, where
the transfer function would be made perfectly flat, meaning that the
efficiency of the rejection would be 100\% for each frequency. The
limitation of the vibration rejection would arise from the
intrinsic noise of the seismometer, which strongly depends on the
frequency. When taking this noise into account, it appears useless
to literally flatten the seismometer response. We numerically
found that a simple first order digital filter compensates the
response function well enough to reach a sensitivity of $5
\times 10^{-9} \rm{g}$ at 1~s, despite the high order of the
internal low pass filter of the seismometer. As for the noise
spectrum given by the manufacturer, we calculated that its
limitation to the interferometer sensitivity would amount to $2
\times 10^{-9} \rm{g}$ at 1~s.

Since the vibration signal is processed with the computer, a
digital filtering seems very favorable. We use a recursive IIR
(Infinite Impulse Response) filter with the following shape: a
unity gain below the lower frequency $f_0$, an increasing slope of
20 dB/decade from $f_0$ to $f_1$, and a constant gain above $f_1$.
We also take benefit of the post-correction process to implement a
non-causal low-pass filter (NCLPF). Such a filtering consists in
processing the sampled data in a forward and backward sense with
respect to time. A positive phase shift induced by the direct
reading will be canceled by the reversed one, whereas the
attenuation is applied twice. In our case, the NCLPF prevents the
IIR filter from amplifying the intrinsic noise of the seismometer
at high frequencies, without affecting the phase advance needed to
improve the rejection. More precisely, the corner frequency of the
NCLPF corresponds to the frequency above which the seismometer
signal doesn't carry any useful information.

After combining the IIR and low-pass non causal filter and before
implementing them during the interferometer measurement, we
checked their effect on the amplitude and the phase shift of the
vibration signal, by exciting the platform again and comparing the
seismometer signal, with and without filter. Excellent agreement
with the expected behavior was found.

\begin{figure}
        \includegraphics[width=13 cm]{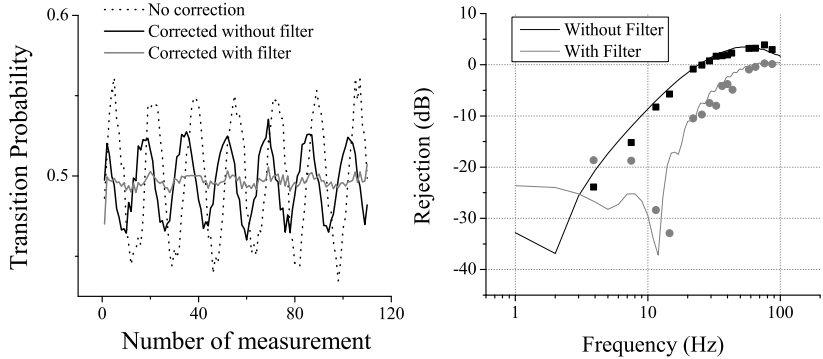}
    \caption{Left: modulation of the transition probability for an excitation frequency of 14 Hz, without correction, with correction, with and without filter. Right: measured rejection efficiencies versus frequency, with and without digital filter. }
    \label{rejection}
\end{figure}

As the atomic signal was also recorded during this measurement,
the influence of the filter on the efficiency of the vibration phase correction could be demonstrated directly on the interferometer signal. To illustrate the gain on the rejection, the modulation of the interferometer signal is displayed in figure \ref{rejection} a) for an excitation frequency of 14 Hz in the different cases : i) we apply no correction, ii) the correction without filter and iii) the correction with the digital filtering. The rejection efficiencies are displayed on figure \ref{rejection} b) as a function of the excitation frequency. They are obtained by calculating the ratio between the amplitudes with and without correction, for the two cases where the digital filter is used or not.

We then implemented the digital filtering in the interferometer phase correction
and operated the interferometer with the nominal vibration noise.
First measurements with the interferometer showed a resolved
influence of the filter, but the rejection was improved by only
15\%. Putting neoprene rubbers below the seismometer legs, the
vibrations above 30 Hz are well damped, so that the signal above
this frequency reaches the seismometer intrinsic noise. This way,
we reduce the contribution of the ``high'' frequencies, for which
it is difficult to well compensate the response of the
seismometer. Nevertheless, the filter does not improve the
rejection efficiency by more than 25\%, whereas the calculation
predicts an improvement by a factor of 3.

\begin{figure}
        \includegraphics[width=9 cm]{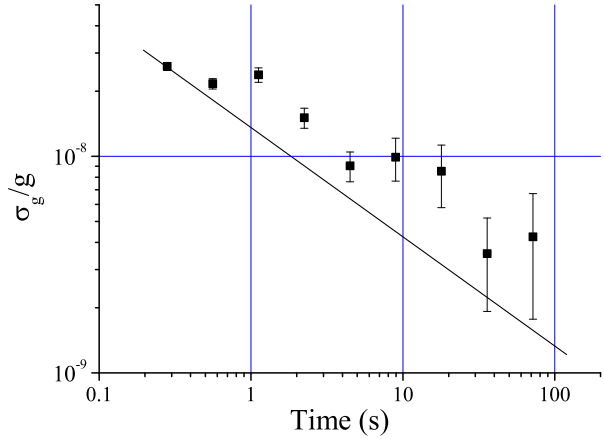}
    \caption{Allan standard deviation of the interferometer phase fluctuations. The sismometer signal is filtered before calculating the phase correction. The air conditioning is switched off.}
    \label{variance}
\end{figure}

At night and with air conditioning switched off, the sensitivity
reaches its best level. Considering the standard deviation at one shot, we deduce an equivalent noise of $1.4 \times 10^{-8} \rm{g}$ at 1 s (see figure \ref{variance}). Deviation from the expected $\tau^{-1/2}$ behavior could be due to the crosstalk with the horizontal directions (see \ref{sismonoisesec}), or to fluctuations of the systematics, most probably to intensity fluctuations. In this situation, with low environmental noise, we find that the filter has no influence, which seems to indicate that the
sensitivity is not limited anymore by vibrational noise.

This sensitivity corresponds to phase fluctuations of 11
mrad/shot, which exceeds the level obtained when summing
(quadratically) all other contributions (4 mrad). As the measurements of
laser phase noise were performed in steady state condition, one
cannot exclude differences in the noise spectra when the
interferometer is operated sequentially, as laser frequencies
undergo abrupt changes and sweeps and Raman lasers are pulsed.
Intrinsic noise of the sismometer, if higher than rated by the
manufacturer, could also be responsible for the observed higher
noise level.

\begin{figure}
       \includegraphics[width=9 cm]{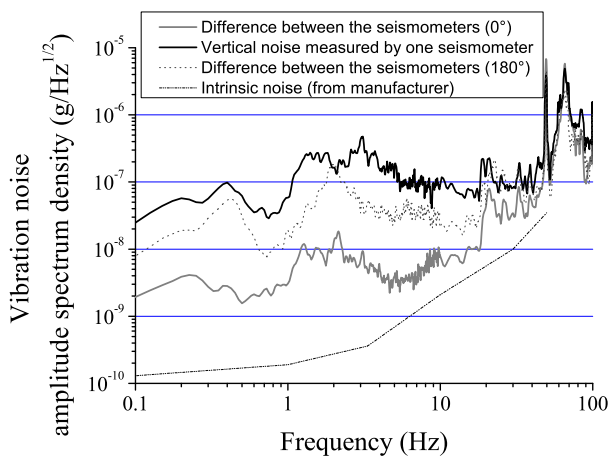}
    \caption{Differential measurements between two sismometers. The grey thick curve displays the difference between seismometers (with identical orientations in the horizontal plane)and the black curve the signal from one of the two sismometers. The dotted curve displays the differential signal for a relative orientation of 180 degrees. The dotted line displays the intrinsic noise of the sensor, as given by the manufacturer.}
    \label{sismonoise}
\end{figure}

\subsubsection{Seismometer intrinsic noise}
\label{sismonoisesec}
First, since coils are used as actuators inside the sismometer, excess noise could be due to magnetic field fluctuations. We therefore measured the sismometer response to magnetic field fluctuations, by modulating the current in a coil placed around it. Having then measured the PSD of ambient magnetic field fluctuations in the laboratory, we finally calculated the equivalent vibration noise, and found less than $1 \times 10^{-9} \rm{g}/\rm{Hz}^{-1/2}$, which rules out magnetic field fluctuations.

We then tried to measure the intrinsic noise of the sismometer, by stacking up two such devices and subtracting their output signals with a low noise differential amplifier. Figure \ref{sismonoise} diplays the result of this differential measurement, as well as the vibration noise measured by the sensors and the intrinsic noise given by the manufacturer.
The rejection efficiency at high frequency (above 10 Hz) is poor, as expected, due to the difference between the transfer functions of the two sensors, which prevents reaching their intrinsic noise. More surprisingly, we find around 2 Hz a poor rejection (only 14 dB) and a broad structure in the spectrum of the differential signal. The output signals from both sismometers being in phase at low frequency and the difference in their scale factor being less than 1\%, the rejection efficiency should reach about 40 dB. The same broad peak also appears in the horizontal acceleration noise spectrum on the platform, about 30 times higher, up to $6\times10^{-7}\rm{g.Hz}^{-1/2}$ at 2 Hz. We thus attribute this residual noise to crosstalk (at the level of a few \%) between horizontal and vertical directions. This assumption was further confirmed by noticing that the rejection was considerably spoiled when positioning the sensors with different orientations in the horizontal plane: see on figure \ref{sismonoise} the differential signal for an angle between the horizontal axes of the two devices of 180 degrees. Still, this parasitic contribution to the vertical acceleration noise at 0 degree is probably partially rejected in the difference, but not completely, due to a difference in the amplitude of the crosstalks, or to the fact that the two devices not being at the same position don't see exactly the same horizontal acceleration. This crosstalk could then constitute the limit in the efficiency of any 1D vibration compensation scheme. A quantitative evaluation of the impact of this effect deserves further studies, and would require the determination of the crosstalk amplitude and transfer function.

\subsection{Other contributions related to transverse displacement}

Many other effects can affect the interferometer phase. In
particular, the most important systematic shifts are related to
Coriolis acceleration and wavefront aberrations.

The Coriolis acceleration leads to a Sagnac effect. If the atoms are released from the molasses with a transverse velocity as low as $100 \mu$m/s, the
shift on the interferometer signal leads to a bias as large as $10^{-9}$g. If
the velocity distribution is symmetric and centered around zero,
then this effect is canceled when detecting the atoms,
provided the detection efficiency is homogeneous across the
whole cloud. Experimental inhomogeneities will in general lead to
a residual shift. We have measured the fluctuations of the mean
velocity of the atomic sample in the horizontal plane by
performing absorption imaging at two different delays after
releasing the atoms. We found short term fluctuations on the order
of $10 \mu$m/s shot. The equivalent noise is 0.03 mrad/shot, which is negligible
with respect to the other effects studied above.

Wavefront aberrations induce an interferometer phase shift that depends on the trajectories of the atoms. A quantitative evaluation of this effect was performed in \cite{Fils05}. Short term fluctuations of the positions (and velocities) of atoms released from a moving molasses were found to limit the sensitivity of their gyro-accelerometer at the level of 0.2 mrad/shot. This limit depends on the details of the wavefront distortions and of the atomic trajectories. It is expected to be different in our geometry, as free falling atoms remain at the center of the laser beam during the interferometer. We have investigated the influence of the velocity fluctuations on the interferometer phase by unbalancing the power in the molasses beams. We found phase shifts of 0.4 mrad per percent change in the intensity ratio, which corresponds to 40 $\mu$m/s velocity change. Velocity fluctuations thus induce phase instability at the level of 0.1 mrad/shot.

\section{Conclusion}

We have extensively studied the different sources of noise in an
atom interferometer, and their influence on the short term
sensitivity of the gravity measurement.

We have demonstrated that a very high sensitivity can be achieved even with a
moderate interrogation time of only 100 ms. This requires an
excellent control of laser phase fluctuations and efficient
detection schemes. We also show that the sensitivity can be
efficiently improved by compensating the phase shifts induced by
vibrations, using the signal of a low noise seismometer, down to a
level limited by intrinsic noise of the sensor and/or by crosstalk between the different measurement axes.

Our final sensitivity is more than twice better than "classical" corner-cube gravimeters. The measurement at the same location, and in the same vibration environment, with FG5\#206 from Institut de Physique du Globe de Strasbourg showed an equivalent sensitivity at 1 s of $4\times10^{-8}$g.

More generally, the work presented here allows to quantify the performances of atom interferometers as a function of interaction time, cycling rate, and sources of perturbations. The same formalism can be used for the design of ultimate sensitivity instruments (such as a space interferometer for instance \cite{Nyman06}), as well as for the realization of lower level compact instruments. In particular, the compensation technique that we have demonstrated in this paper is particularly attractive for the development of a simple and compact instrument, which could reach high sensitivities without vibration isolation.

{\bf Acknowledgments}

We would like to thank Patrick Cheinet for his contribution
in early stage of the experiment, David Holleville for his help in the mechanics and optics design, the Institut Francilien pour la Recherche sur les Atomes Froids (IFRAF) and the European
Union (FINAQS) for financial support. J.L.G.
thanks DGA for supporting his work.


\begin{thebibliography}{10}

\bibitem{Borde89} Ch. J. Bord\'{e}, Physics Letters A \textbf{140}, 10 (1989).


\bibitem{Niebauer95} T.M. Niebauer, G.S. Sasagawa, J.E. Faller, R. Hilt, F.
Klopping, Metrologia \textbf{32}, 159 (1995).

\bibitem{Kasevich91}  M. Kasevich and S. Chu, Phys. Rev. Lett. \textbf{67}, 181 (1991)

\bibitem{Riehle91} F. Riehle, Th. Kisters, A. Witte, and J. Helmcke, Ch. J. Bordé, Phys. Rev. Lett. \textbf{67}, 177 (1991).

\bibitem{Canuel06} B. Canuel, F. Leduc, D. Holleville, A. Gauguet, J. Fils, A. Virdis, A. Clairon, N. Dimarcq, Ch. J. Bordé, A. Landragin, and P. Bouyer, Phys. Rev. Lett. \textbf{97}, 010402 (2006).

\bibitem{Peters01} A. Peters, K. Y. Chung, S. Chu, Metrologia \textbf{38}, 25 (2001), H. Muller, Arxiv (2007),  H. Mueller, S. Chiow, S. Herrmann, S. Chu, K. Y. Chung, arXiv:0710.3768[gr-qc]


\bibitem{Gustavson02} T. L. Gustavson, A. Landragin, and M. A. Kasevich, Classical Quantum Gravity \textbf{17}, 2385 (2000).

\bibitem{Bertoldi06} A. Bertoldi, G. Lamporesi, L. Cacciapuoti, M. de Angelis, M. Fattori, T. Petelski, A. Peters, M. Prevedelli, J. Stuhler and G.M. Tino, Eur. Phys. J. D \textbf{40}, 271 (2006).

\bibitem{Fixler07} J. B. Fixler, G. T. Foster, J. M. McGuirk, M. A. Kasevich, Science Magazine Vol. 315. no. 5808, 74 (2007).

\bibitem{APB} Special Issue: "Quantum Mechanics for Space Application: From Quantum Optics to Atom Optics and General Relativity"
 Appl. Phys. B \textbf{84} (2006).

\bibitem{Geneves05} G. Genev\`{e}s et al., IEEE Trans. on Instr. and Meas. \textbf{54}, 850 (2005).


\bibitem{Steiner07} R. L. Steiner, E. R. Williams, R. Liu, and D. B. Newell, IEEE Trans. on Instrum. Meas. \textbf{56}, 592 (2005).

\bibitem{Robinson07} I. Robinson and B. P. Kibble, Metrologia \textbf{44}, 427 (2007).


\bibitem{Yver03} F. Yver-Leduc, P. Cheinet, J. Fils, A. Clairon, N. Dimarcq,
D. Holleville, P. Bouyer, A. Landragin, J. Opt. B : Quantum
Semiclas. Optics \textbf{5}, S136 (2003).

\bibitem{Cheinet06} P. Cheinet, F. Pereira Dos Santos, T. Petelski, J. Le Gou\"{e}t, J. Kim, K.T. Therkildsen, A. Clairon and A. Landragin, Appl. Phys. B \textbf{84}, 643 (2006).

\bibitem{Borde01} Ch. J. Bord\'{e}, C.R. Acad. Sci. Paris, t.2, Série IV, 509-530 (2001).

\bibitem{Baillard06} X. Baillard, A. Gauguet, S. Bize, P. Lemonde, Ph. Laurent, A. Clairon and P. Rosenbusch, Optics Communications \textbf{266}, 609 (2006).


\bibitem{Cheinet05} P. Cheinet, B. Canuel, F. Pereira Dos Santos, A. Gauguet, F. Leduc, A. Landragin, accepted for publication in IEEE Trans. on Instrum. Meas., Arxiv physics/0510197 (2005).

\bibitem{Dick87} G. J. Dick, "Local Ocillator induced
instabilities," in Proc. Nineteenth Annual Precise Time and Time
Interval, pp. 133-147 (1987); G. Santarelli, C. Audoin, A. Makdissi, P. Laurent
, G.J. Dick and A. Clairon, IEEE Trans. Ultrason. Ferroelect. Control \textbf{45}, 887 (1998).

\bibitem{Nyman06} R.A. Nyman, G. Varoquaux, F. Lienhart, D. Chambon, S. Boussen, J.-F. Clément, T. Müller,
G. Santarelli, F. Pereira Dos Santos, A. Clairon, A. Bresson, A.
Landragin et P. Bouyer, Appl. Phys. B \textbf{84}, 673 (2006).

\bibitem{Santarelli94} G. Santarelli, A. Clairon, S.N. Lea and G. Tino, Opt. Comm. \textbf{104}, 339 (1994).

\bibitem{LeGouet07} J. Le Gou\"{e}t, P. Cheinet, J. Kim, D. Holleville, A. Clairon, A. Landragin, and F. Pereira Dos Santos, Eur. Phys. J. D \textbf{44}, 419 (2007).

\bibitem{Weiss94} D. S. Weiss, B. C. Young et S. Chu, Appl. Phys. B \textbf{59}, 217 (1994).

\bibitem{TheseBen} Benjamin Canuel, http://tel.archives-ouvertes.fr/tel-00193288/fr/

\bibitem{Clade} P. Clad\'{e}, E. de Mirandes, M. Cadoret, S. Guellati-Khélifa, C. Schwob, F. Nez, L. Julien, and F. Biraben, Phys. Rev. A \textbf{74}, 052109 (2006).


\bibitem{Itano93} W. M. Itano, J. C. Bergquist, J. J. Bollinger, J. M. Gilligan, D. J. Heinzen, F. L. Moore, M. G. Raizen, and D. J. Wineland, Phys. Rev. A \textbf{47}, 3554 (1993).

\bibitem{Santarelli99} G. Santarelli, Ph. Laurent, P.
Lemonde, A. Clairon, A. G. Mann, S. Chang, and A. N. Luiten, C.
Salomon, Phys. Rev. Lett. \textbf{82}, 4619 (1999).

\bibitem{McGuirk01} J. M. McGuirk, G. T. Foster, J. B. Fixler, and M. A. Kasevich, Opt. Lett. \textbf{26}, 364 (2001).

\bibitem{McGuirk02} J. M. McGuirk, G. T. Foster, J. B. Fixler, M. J. Snadden, and M. A. Kasevich, Phys. Rev. A \textbf{65}, 033608 (2002).

\bibitem{Fils05}  J. Fils, F. Leduc, P. Bouyer, D. Holleville, N. Dimarcq, A. Clairon et A. Landragin, Eur. Phys. J. D \textbf{36}, 257 (2005).

\end{thebibliography}

\end{document}